\documentclass[conference]{IEEEtran}

\IEEEoverridecommandlockouts

\usepackage[utf8]{inputenc} 
\usepackage[T1]{fontenc}    
\usepackage{url} 
\usepackage{amsmath}
\usepackage{booktabs}       
\usepackage{amsfonts}       
\usepackage{nicefrac}       
\usepackage{microtype}      
\usepackage{graphicx}
\usepackage{gensymb}
\usepackage{subfigure}
\usepackage{xcolor}

\usepackage[utf8]{inputenc}
\usepackage{tcolorbox}
\tcbuselibrary{skins} 
\usepackage{listings}
\lstdefinelanguage{json}{
    basicstyle=\normalfont\ttfamily,
    numbers=left,
    numberstyle=\scriptsize,
    stepnumber=1,
    numbersep=8pt,
    showstringspaces=false,
    breaklines=true,
    frame=lines,
    backgroundcolor=\color{white},
    stringstyle=\color{blue},
    commentstyle=\color{gray},
    keywordstyle=\color{red},
    ndkeywordstyle=\color{red}
}
\usepackage{algorithm}
\usepackage{algorithmic}
\usepackage{cite}
\usepackage[symbol]{footmisc}

\definecolor{color_CPS}{rgb}{0, 0, 0}
\newcommand{\hanqing}[1]{{\color{color_CPS} #1}}

\definecolor{myblue}{rgb}{0,0,0}
\newcommand{\hq}[1]{{\color{myblue} #1}}
\definecolor{brown}{rgb}{0.0, 0.0, 0.0}
\definecolor{blue}{rgb}{0.0, 0.0, 0.0}
\definecolor{violet}{rgb}{0.0, 0.0, 0.0}
\definecolor{purple}{rgb}{0.0, 0.0, 0.0}
\definecolor{red}{rgb}{0.0, 0.0, 0.0}
\definecolor{teal}{rgb}{0, 0, 0}
\def\algoname{AitL-RL}
\def\hitl{HITL}

\usepackage{hyperref}
\hypersetup{
    colorlinks=true,
    linkcolor=magenta,
    filecolor=magenta,      
    urlcolor=magenta,
    pdftitle={Overleaf Example},
    pdfpagemode=FullScreen,
    }

\title{An LLM-Based Digital Twin \\ for Optimizing Human-in-the Loop Systems}

\author{
\IEEEauthorblockN{Hanqing~Yang\IEEEauthorrefmark{1},
Marie~Siew\IEEEauthorrefmark{2}, and
Carlee~Joe-Wong\IEEEauthorrefmark{1}
}
\IEEEauthorblockA{\IEEEauthorrefmark{1}Electrical and Computer Engineering,
Carnegie Mellon University, Pittsburgh, PA 15213 USA}
\IEEEauthorblockA{\IEEEauthorrefmark{2}Information Systems Technology and Design Pillar, Singapore University of Technology and Design, 487372 Singapore}
\IEEEauthorblockA{Emails: \{hanqing3, cjoewong\}@andrew.cmu.edu, marie\_siew@sutd.edu.sg 
}
\thanks{GitHub repository: \url{https://github.com/HappyEureka/LLM_digital_twin}.\newline
\indent\indent We gratefully acknowledge support from Microsoft's Accelerating Foundation Models Research initiative.
M. Siew is supported by the Faculty Early Career Award at SUTD. (Corresponding author: Marie Siew).
}}
\AtBeginEnvironment{tcolorbox}{\small}
\begin{document}
\maketitle
\begin{abstract}
The increasing prevalence of \hanqing{Cyber-Physical Systems and the Internet of Things (CPS-IoT) applications and Foundation Models} are enabling new applications that leverage real-time control of the environment. For example, real-time control of Heating, Ventilation and Air-Conditioning (HVAC) systems can reduce its usage when not needed for the comfort of human occupants, hence reducing energy consumption. Collecting real-time feedback on human preferences in such human-in-the-loop (\hitl) systems, however, is difficult in practice. We propose the use of large language models (LLMs) to deal with the challenges of dynamic environments and difficult-to-obtain data in CPS optimization. 
\hanqing{
In this paper, we present a case study that employs LLM agents to mimic the behaviors and thermal preferences of various population groups (e.g. young families, the elderly) in a shopping mall. 
The aggregated thermal preferences are integrated into an agent-in-the-loop based reinforcement learning algorithm \algoname, which employs the LLM as a dynamic simulation of the physical environment to learn how to balance between energy savings and occupant comfort.} 
Our results show that LLMs are capable of simulating complex population movements within large open spaces.
Besides, \algoname demonstrates superior performance 
compared to the popular existing policy of set point control, suggesting that adaptive and personalized decision-making is critical for efficient optimization \hanqing{in CPS-IoT applications. Through this case study, we demonstrate the potential of integrating advanced Foundation Models like LLMs into CPS-IoT to enhance system adaptability and efficiency.} The project's code can be found on our
\href{https://github.com/HappyEureka/LLM_digital_twin}{GitHub repository}.
\end{abstract}

\section{Introduction}
Large language models (LLMs) have been drawing significant attention, ever since the announcement of ChatGPT in November 2022~\cite{chatgpt}. As LLMs have demonstrated powerful language understanding, reasoning, and generation capabilities, there has been a growing research area \cite{wang2023survey,park2023generative,lan2023llm, xu2023exploring, mao2023alympics} involving generating multiple \textbf{LLM-powered agents} to simulate human behavior, interaction, and decision making, 
for various tasks across domains: 
to play strategic communication games like Avalon \cite{lan2023llm} and Werewolf \cite{xu2023exploring}, to participate in auctions \cite{mao2023alympics}, to see if they can model human behavior in classic social science experiments \cite{aher2023using}, 
for planning in industrial automation via digital twins \cite{xia2023towards}, etc. 
In many of the above works, LLM-powered agents were given different profiles, and prompted to take the role of their profile in their interactions.
In this paper, we propose using LLM-powered agents to help build \textbf{digital twins for \textcolor{red}{adaptive control of} human-in-the-loop (\hitl) systems}. \textcolor{red}{While technological advances have enabled the implementation of such adaptive control systems for applications like urban transportation planning and human robot collaboration in manufacturing, collecting real-time data on human preferences remains a critical challenge.}
LLM-powered agents can \textcolor{red}{solve this challenge by modeling} how multiple population types interact with the system (e.g. interests, mobility patterns, preferences, etc).
Besides this, the LLM-powered autonomous agents possess \textit{memory and planning capabilities}, which helps to model human decision making.
\textcolor{red}{We can then use data generated by the LLM to learn how a system should adapt to its human users.}
In this paper, we \textcolor{purple}{develop an LLM-powered digital twin to simulate user behavior in a building, and apply it to temperature control.}
\textcolor{teal}{Similar setups can be used for other human in the loop system control settings.}


\hanqing{As Cyber-Physical Systems and the Internet of Things (CPS-IoT) applications integrate into our daily lives, they unlock new possibilities such as \hitl~control systems. One \textbf{challenge} for such systems lies in accurately representing the complex dynamics of diverse groups' interactions with their environments  \cite{ma2018data}. Our solution uses a LLM to generate detailed simulations of environments such as a mall, producing data on human-environment interactions without privacy concerns. 
Furthermore, multi-objective optimization presents a significant challenge in the absence of a closed-form model to capture human feedback \cite{cui2017multi}. To address this, we employ Reinforcement Learning (RL), allowing for the dynamic optimization of policies based on real-time human feedback iteratively.}

 
%
\textbf{Our case study:} The building sector consumes around $36\%$ of global energy \cite{santamouris2021present}, with Heating, Ventilation and Air-Conditioning (HVAC) systems the largest contributors \cite{katili2015space}.
In public buildings, cooling can account for more than $50\%$ of energy usage \cite{katili2015space}. 
Temperature control solutions 
can potentially reduce energy usage by adjusting the HVAC, as long as they adaptively balance energy usage with occupant comfort.


We consider a highly occupied public space, such as a public mall, a cinema, a library, or a communal office space. 
Unlike homes or offices with individual temperature control, 
public spaces typically lack user-adjustable settings.
Nevertheless, users may find the temperature too cold. At the same time, increasing the temperature could also bring about energy savings.
\textcolor{violet}{One way to take advantage of these opportunities is a \hitl~system for temperature control, which collects real-time user feedback, e.g., through a mobile app, on their comfort levels, 
\textcolor{brown}{and learns to adjust temperatures so as to} balance energy savings with user-comfort.  
Nevertheless, gathering crowd-sourced user feedback can be challenging. Users may be too busy (e.g., being immersed when shopping or watching movies), or may not take the application seriously, giving false data on their preferences. 
}
%
\textcolor{violet}{Therefore, we propose a LLM-based digital twin solution, 
which uses LLM agents to generate data for training a RL algorithm that 
balances energy savings with user comfort. 
Our \textbf{contributions} are summarized as follows:} 
\hq{
\begin{itemize}
    \item \textcolor{violet}{
    We use GPT to simulate a building (mall) with multiple stores.
    We use \textbf{LLM-powered autonomous agents}, each representing a population group, and imbued with the group's unique characteristics like mobility patterns 
    and temperature preferences e.g., families with young children, elderly couples, etc.} 
    \textcolor{violet}{
    In every iteration, GPT 3.5 and 4 are used to simulate each population group's arrivals and distribution across stores. Note that only querying is performed, maintaining a low computational footprint. 
    }
    \item We train the RL algorithm offline, using data generated by the digital twin, to optimize energy savings and user comfort.
    We propose both \textbf{centralized and distributed temperature control versions} of the algorithm and show that the policies trained by the algorithms can generalize to new data when deployed in online scenarios.
    \item Our \textbf{experimental results} demonstrate the superior performance of distributed control settings over centralized control and set-point control settings.
\end{itemize}
}

\textcolor{brown}{We \textbf{release the simulator} for other researchers, industry partners and the public to use and 
customize for specific buildings; 
similar LLM-based digital twins can be used for other \hitl~systems in domains like transportation, urban and retail planning. We present our agents-in-the-loop solution in Section~\ref{sec:aitl} and our evaluation results in Section~\ref{sec:results} before concluding in Section~\ref{sec:conclusion}.
}


\section{Agent-in-the-Loop Learning 
}\label{sec:aitl}
Optimizing distributed \hitl~systems for conflicting objectives, e.g., in our context of balancing energy consumption with utility, is challenging. 
Such systems are often inherently complex and lack real-life data, which is often 
\textcolor{violet}{difficult to obtain as the system involves human activity, interaction and preferences}. Fortunately, advancements in \hanqing{Foundation Models} have opened new possibilities for simulating complex environments with believable human behaviours \textcolor{violet}{\cite{wang2023survey,park2023generative,lan2023llm, xu2023exploring, mao2023alympics}}. By leveraging these models, it is feasible to create customized simulations on which we can train reinforcement learning (RL) techniques for the optimization of distributed systems. This method bypasses the limitations of the traditional data collection process, offering a cost-effective and time-efficient methodology for adaptive distributed systems optimization.

In this study, we introduce a novel framework for distributed system control optimization, the LLM-aided agents-in-the-loop reinforcement learning framework (\algoname). We focus on \textcolor{violet}{ temperature optimization for air-conditioners in} large public spaces, while balancing energy savings and user comfort. 
We study two control settings for HVAC systems: \textbf{centralized} and \textbf{distributed}. In the centralized setting, the entire building is maintained at a uniform temperature. In contrast, the distributed setting involves multiple control points across subsections of the mall, 
as users in different parts of the building may have varying thermal comfort requirements e.g. due to usage-based reasons (libraries vs cinemas) or location-based reasons (some locations are nearer the entrance).

\subsection{Environment Setup}
\begin{figure}[t]
    \begin{center}
        \includegraphics[angle=0,scale=0.23]{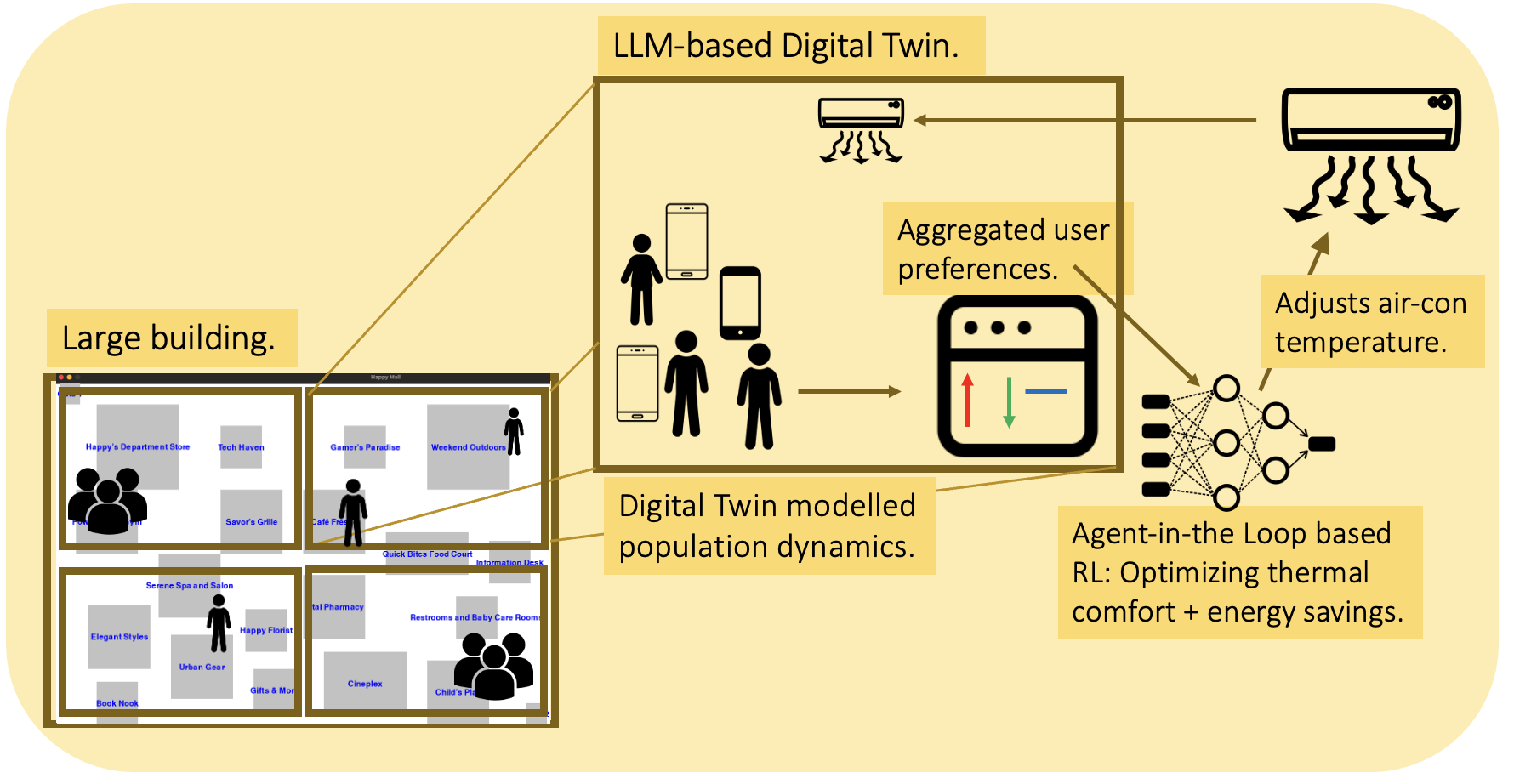}
        \caption{The LLM-based Digital Twin Agent in the Loop Distributed Control (\algoname) Pipeline. \textcolor{violet}{The LLM-based digital twin simulates population behavior in the mall across the day, with multiple population groups such as "teen shoppers". Based on the simulation, user preferences are aggregated and input into the Agent-in-the-loop RL algorithm for offline training 
        to optimize user comfort and energy savings. }}
        \label{fig:systemOverview}
    \end{center}
    \vspace{-6pt}
\end{figure}

With the aid of LLMs, we simulate 
an environment that mirrors a typical day 
in a shopping mall. Different population groups and their activities in this space are modelled, with updates at each designated checkpoint. The environment consists of three components: open space, population groups, and movements of population groups in the open space. Fig. \ref{fig:systemOverview} shows the simulation system.

\begin{figure}[b]
    \centering
        \includegraphics[width=.38\textwidth]{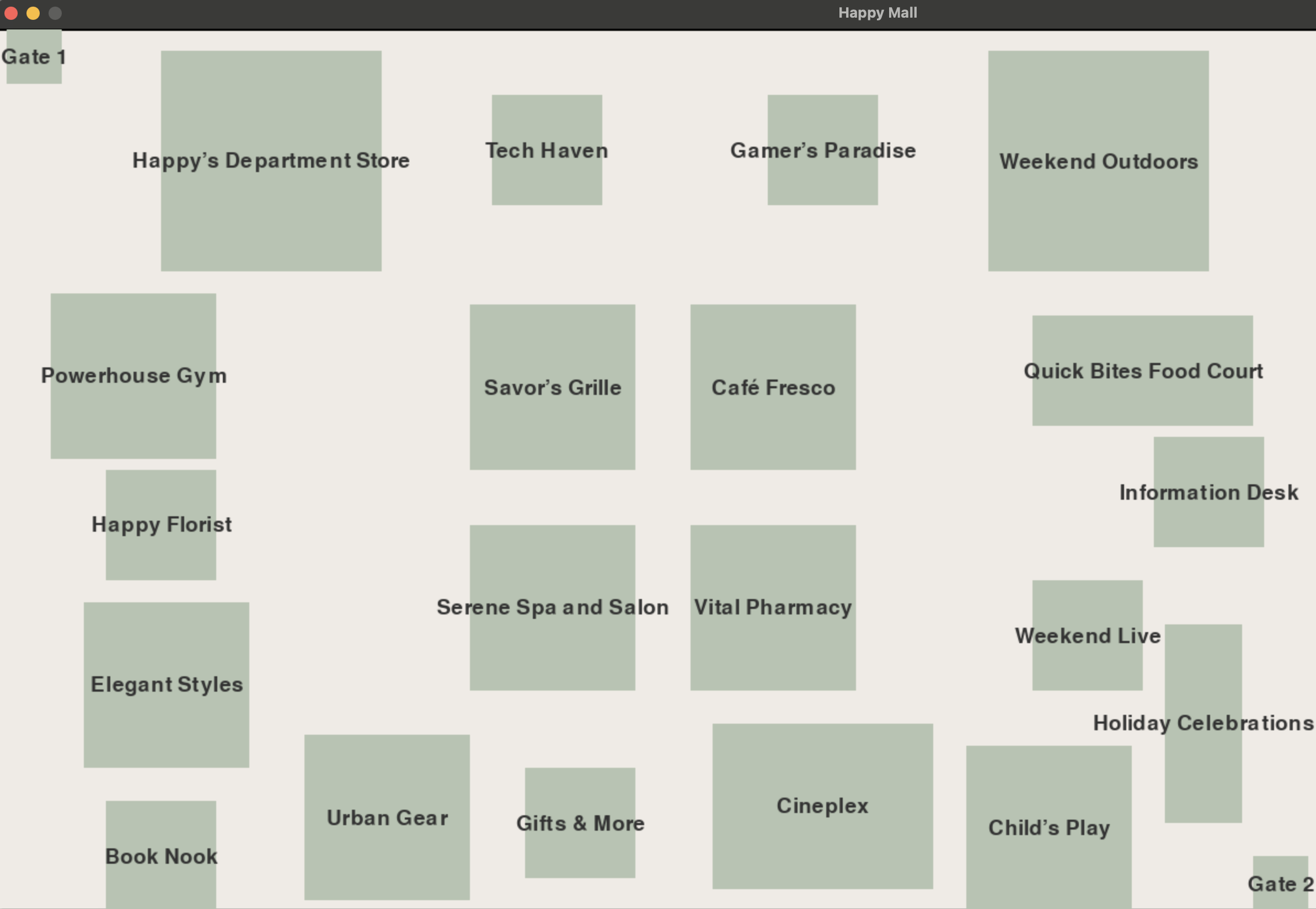}
    \caption{The \textcolor{violet}{layout of the shopping mall named Happy Mall, whose floor plan was generated by GPT 4}, after minor manual adjustments to better replicate a real mall setting.}
    \label{fig:simResults}
    \vspace{-5pt}
\end{figure}
\hanqing{Although it is possible to use a real-world mall setting, at this preliminary stage, LLM simulations offer a simplified and controllable setting that allows for focused study on specific aspects of CPS devices' interactions with humans. Additionally, simulations provide scalability and reproducibility across various conditions and scenarios, facilitating rapid prototyping and iteration.}

\textbf{Open space}: An open space resembling a large shopping mall, catering to approximately 3000 daily consumers, operates from 10 a.m. to 8 p.m. The mall's design is generated using ChatGPT 4 (refer to Fig. 2), starting with an initial prompt and refined iteratively. Minor manual adjustments were made to store locations for a more realistic layout.
\begin{tcolorbox}
    \textbf{Initial generation}: Please simulate a mid-size shopping mall named Happy Mall in JSON format. Further, please create a map of the Happy Mall on a 1600 by 1200 canvas, indicating the location of each component as described.\\
    \textbf{Refinement}: Please make sure each component fits within the border. In addition, adjust the store distribution to vary in size and spacing across the canvas more realistically, like in a real mall setting. Please make corrections.
\end{tcolorbox}

\textbf{Population groups}: The individuals in the open space are simulated by categorizing them into groups, such as Teen Shoppers, Elderly Couples, Tourist Groups, etc., achieved through prompts to ChatGPT 4 with the category prompt. The movements of population groups are modelled using ChatGPT 3.5 at each 30-minute interval. The population and distribution prompts determine the number of individuals per group and the distribution of the groups across stores, respectively.

\begin{tcolorbox}
    \textbf{Category prompt}: In the context of the Happy Mall, identify the different visitor groups, focusing on their occupation and age. Please provide this information in JSON format, including a group description, thermal preference, and their comfortable temperature range in degrees Celsius, divided into 'high' and 'low' fields.\\
    \textbf{Population prompt}: Located in a town center with a population of \{TOWN POPULATION\}, the Happy Mall opens at \{OPEN TIME\} and closes at \{CLOSE TIME\}. Currently, it's \{CURRENT TIME\}. Descriptions of different groups in the mall are as follows: \{DESCRIPTIONS OF GROUPS\}. Given the time of day, list the number of people in each group using the format: [group name]: [number]; [reason].\\
    \textbf{Distribution prompt}: Currently, it's \{CURRENT TIME\} and \{INDOOR TEMPERATURE\} \degree C. The descriptions of the stores in the Happy Mall are as follows. \{DESCRIPTIONS OF STORES\}. You belong to \{GROUP NAME\}, described as \{DESCRIPTION OF THE GROUP\}, with a thermal preference of \{THERMAL PREFERENCE OF THE GROUP\}. What is your group's distribution in the mall? List the percentile for each store using the following format. [store]: [percentile]; [reason].
\end{tcolorbox}

\begin{figure}[t]
    \centering
        \subfigure[\hanqing{The average population distribution changes over time, with different lines highlighting distinct movement patterns for various groups.}]{\includegraphics[width=.41\textwidth]{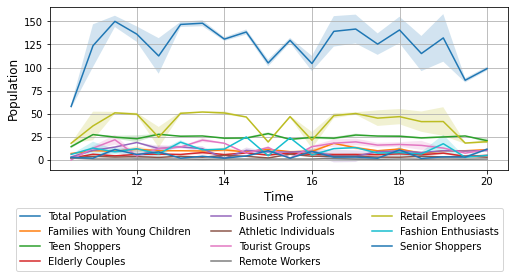}}
        \subfigure[\hanqing{The average population distribution change for each group in a day.} The changes in colour gradient demonstrate the density for each group throughout the day.]{\includegraphics[width=.41\textwidth]{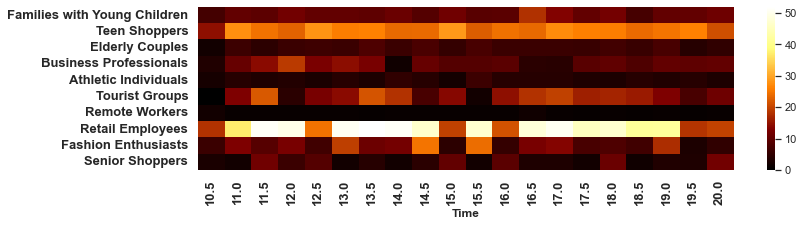}}
        \subfigure[\hanqing{The average population distribution change for each store in a day.} The changes in colour gradient represent the frequency of store visits.]{\includegraphics[width=.41\textwidth]{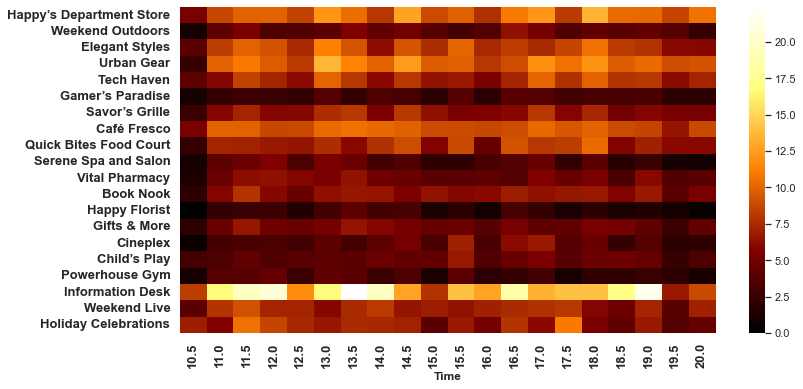}}
    \caption{Simulation of movement in Happy Mall over a day. \textcolor{violet}{These distributions for the digital twin are generated by GPT 3.5 in each time slot.} The population groups exhibit different temporal patterns (results averaged over 10 experiments).}
    \label{fig:popDistribution} [-2ex]
\end{figure}

The average change in population distribution for each group is shown in Fig \ref{fig:popDistribution}.a and Fig \ref{fig:popDistribution}.b. In Fig \ref{fig:popDistribution}.a, the line plot shows the average population counts for different groups within the mall, along with the confidence level from 10 experiments. Each line represents the average change in the population of a specific group in the mall, highlighting various peak times such as noon, afternoon, and evening. Fig \ref{fig:popDistribution}.b is a heat map that shows the detail of the change in population for each group. The colour gradient indicates the density of each group in the mall at each checkpoint, where brighter shades represent higher density. We can clearly see that different population groups have different movement patterns, such as families with young children peaking at 2 pm and 4:30 pm, teen shoppers peaking at 12:30 pm and 3 pm, etc. Fig \ref{fig:popDistribution}.c presents a heat map showing the average population changes in each store. Each store demonstrates unique peak hours.

\textbf{Non-deterministic responses:} \hanqing{A critical challenge with LLMs is their non-deterministic responses. However, our experiments demonstrate that by setting the temperature of the LLMs to zero, the simulations exhibit consistent trends with tolerable variations. 
Human behaviors themselves are also inherently variable and context-dependent. People respond differently based on a multitude of factors such as mood and personal background. This variability is closely mirrored by the non-deterministic nature of LLMs, which makes LLMs particularly promising for simulating real human behavior, capturing the essence of human unpredictability and diversity in responses.
In future works, the LLMs prompts for generating the digital twin simulations can be augmented with the limited existing datasets, through retrieval augmented generation.} 



\subsection{Reinforcement Learning}
The simulation provides information on the movements of each group across stores. The information \textcolor{violet}{which contains the thermal preferences of the different groups} is fed into the \algoname~framework to optimize the HVAC system for energy efficiency while maintaining comfort for shoppers throughout the day in both centralized and distributed control settings. We train \algoname~offline in an environment with the following characteristics:

\textbf{State space}: The state space is an amalgamation of the \textit{human input} [number of votes for "temperature increase", number of votes for "temperature decrease", \textcolor{violet}{number of votes for "constant temperature"}], along with the \textcolor{violet}{external environment} variables [current time, outdoor temperature, indoor temperature], as well as the group occupancy for each space.

We simulate the human input with our LLM-powered agents.
At each checkpoint, we collect user feedback \textcolor{violet}{from the different population groups, i.e. the LLM-powered agents on their perceptions of the current indoor temperature. 
These perceptions are based on the thermal comfort ranges for each population group, generated while creating the environment.}
 Users can either vote 
 \textcolor{violet}{for a temperature increase, decrease,} or no change, signifying feelings of cold, heat, or comfort respectively. 

\textbf{Action}: The RL algorithm controls the HVAC system, which controls the temperature of the indoor environment. For safety reasons, the actions (temperature set) range between $17$ to $29 \degree C$, with intervals of $0.5 \degree C$. In the centralized control setting, one RL model is trained to control the temperature for the entire mall; in the distributed control setting, multiple RL models are trained, each responsible for the temperature of an individual store. As a result, the centralized approach maintains a uniform temperature throughout the mall, whereas the distributed approach allows for personalized control in each store, \textcolor{violet}{or subsection of the building.}

\textbf{Transition}: A transition happens at each checkpoint, driven by the arrivals, departures, and mobility patterns of the population and the collected thermal feedback.
\textcolor{violet}{Every timestamp, the collected feedback is input into our RL algorithm, \algoname, and the algorithm determines the temperature for the next timestamp.}
The environment transitions from one timestamp (checkpoint) to the next timestamp. \textcolor{red}{Note that the thermal feedback is a function of the previous timestep's action, which is the indoor mall temperature.}

\textbf{Reward}: Our objective is to optimize for both energy cost and user comfort by maximizing the reward function: 
\begin{equation}
   \textit{Reward}=w_c \times \textit{UserComfort}(t) - w_e \times \textit{EnergyUsage}(t),
\end{equation}
which is a weighted sum of user thermal comfort ($\textit{UserComfort}(t)$) and energy savings ($\textit{EnergyUsage}(t)$), with $w_c$ and $w_e$ representing the respective weights. By adjusting these weights, the algorithm can shift its focus, emphasizing one aspect over the other.

\textcolor{red}{
The \textit{user comfort} score is determined by subtracting the number of users experiencing discomfort (cold or heat) multiplied by a weight of 1, and adding the number of comfortable users multiplied by a weight of 2. 
}
\textcolor{red}{
We model \textit{energy usage} based on the heat transfer equation \cite{holman1986heat}:
\begin{equation}
    \textit{EnergyUsage}= \frac{m c_a \Delta T}{\textit{EER}},
\end{equation}
where $m$ is the mass of the room, $c_a$ is the heat capacity of dry air, $\Delta T$ is the temperature difference between the outdoor (ambient) temperature and the indoor (air-con controlled) temperature, and \textit{EER} is the ratio of the cooling capacity to the power input.
The area of the space is set as $4890$ $m^2$, and its height is set as $3$ meters \cite{shen2019}. The density of dry air is $1.275$ $kg/L$. With this, we compute the mass as $\text{volume} \times \text{density}$. The heat capacity of dry air is $1.00$ $J/ (g K)$. \hq{The ambient temperature is set at 30 $\degree C$.}
}

To optimize the reward function, 
we input the state space into a deep Q-learning algorithm \cite{mnih2015human}. 
The output of the neural network is the \textbf{action} taken. In Q-learning, the Q-function $Q(s, a)$ represents the value (sum of expected discounted reward) of taking action $a$ at state $s$, and the algorithm's aim is to iterate until convergence at the true Q-value.
\begin{equation}
    Q(s,a) \gets (1-\alpha) Q(s,a)+\alpha (Reward(t) + \gamma \max_a Q(s',a) ),
\end{equation}
where $\alpha$ is the learning rate, and $\gamma$ is the discount factor.
In deep Q-learning, the neural network parameterizes the Q-function.
The loss function which the neural network optimizes is
\begin{equation}
    L=[Reward+ \gamma \max_{a'} Q(s',a';\theta') - Q(s,a;\theta)]^2,
\end{equation}

We set the following parameters for the RL algorithm within the \algoname\ framework (Table 1).
\\[-4ex]
\begin{table}[!htbp]
  \caption{Hyper-parameters for AiTL-RL}
  \label{tab:hyperparameter_table}
  \centering
  \begin{tabular}{lclc}
    \toprule
    Name & Value & Name & Value \\
    \midrule
    Batch size & 128 & Learning Rate & $10^{-4}$ \\
    Discount $\gamma$ & 0.99 & $w_e$ & $1/220$ \\
    Init. $\epsilon$ & 0.9 & $w_c$ & 2.2 \\
    Final $\epsilon$ & 0.05 & $\tau$ & $5 \times 10^{-3}$ \\
    $\epsilon$ decay & 2000 & Exploration rate & $0.85e^{-\frac{x}{2000}}$\\
    \bottomrule
  \end{tabular}
  \\[-1.0ex]
\end{table}
\section{Evaluation and Results}\label{sec:results}
We trained \algoname\ offline using data generated by the LLM-powered digital twin in both centralized and distributed settings. For comparison, we used the set-point control approach as a baseline, and we trained \algoname with a specific focus on either user comfort or energy cost as alternative baselines. 
\subsection{Experiment Design}
Our objective is to train an algorithm to learn the optimal policy for temperature control in a mall, balancing user comfort and energy costs. We have designed five different training settings for the algorithm and evaluated their performances. These include \algoname\ in both \textbf{Centralized Control} and \textbf{Distributed Control} settings. We compare their results to those of three baseline policies:

\textbf{Set-point Control}: Set-point Control refers to the technique of Variable Refrigerant Flow (VRF) systems \cite{kim2020energy}, as VRF systems are often used in large buildings. Following the approach of \cite{kim2020energy}, we use a constant temperature of $25\degree C$, as ASHRAE 55-2017 \cite{ashrae2017standard} recommends a summertime thermal comfort zone of $23 - 26 \degree C$, and $25\degree C$ has been found to yield the highest reward in trials.

\textbf{User-comfort-focused Control}: the algorithm's training will focus entirely on user comfort by setting $w_e$ to 0.

\textbf{Energy-focused Control}: In this setting, the algorithm's training will focus entirely on energy cost by setting $w_c$ to 0.

\subsection{Reinforcement Learning Performance}
\begin{figure}[h]
    \centering
        {\includegraphics[width=.35\textwidth]{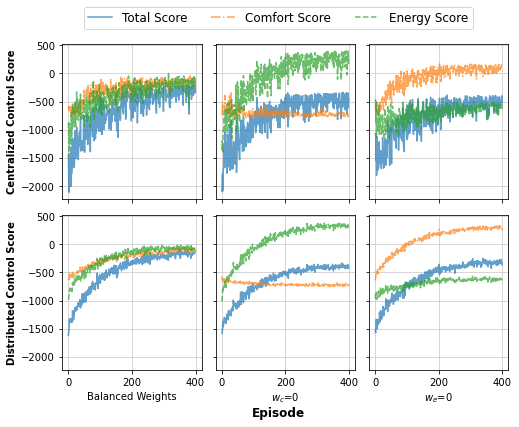}}
    \caption{Comparison of training convergence using different methods: balanced weights (considering both user comfort and energy cost), energy-focused, and user-comfort-focused. All settings converge, and the balanced weights approach achieves the best overall score, emphasizing the need for equal consideration of multiple aspects.\\[-3ex]}
    \label{fig:convergePlots}
\end{figure}

We employed reinforcement learning to determine the optimal temperature control policy. Figure \ref{fig:convergePlots} illustrates the training process for both centralized and distributed settings. Each episode represents a day-long simulation with scores calculated at every half-hour checkpoint. The generated hourly population distribution was used for efficient offline training.
The converged RL network was then applied for an online scenario distinct from the training data.
At each checkpoint, the rewards were re-calculated based on the policy proposed by the reinforcement learning algorithm.

\textbf{Performance metrics:} The scores represent the total daily cumulative rewards, with \textbf{comfort score} reflecting the daily cumulative comfort reward and \textbf{energy score} reflecting the daily cumulative energy reward. The \textbf{total score} represents the weighted comfort score and the energy score. The first column demonstrates training with balanced weights, indicating the algorithm learns the optimal policy by considering both user comfort and energy cost. The second column represents training in the energy-focused control setting, disregarding user comfort. The right column displays training in the user-comfort-focused control setting, ignoring energy costs. The figure reveals that the balanced approach yields the highest total score. While baseline policies can effectively optimize for either energy or comfort, such a singular focus adversely impacts the other component, leading to a lower overall score.

\begin{figure}[h]
    \centering
        {\includegraphics[width=.38\textwidth]{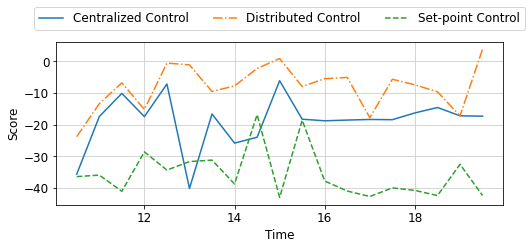}}
        \\[-2ex]
    \caption{Comparison of policy performance in an online scenario using a balanced weights approach. The \algoname\ policies outperform the set-point policy, and distributed control is more efficient than centralized control.} 
    \label{fig:rewardCompare}
\end{figure}

\textbf{\algoname~results:} Figure \ref{fig:rewardCompare} shows the changes in total scores, which combine weighted user comfort and energy cost scores, under centralized, distributed, and set-point control policies using the balanced weights method. The results are observed on newly generated days in an online scenario distinct from the training data, which incorporates real-time control of the system and feedback from the population. \textit{\algoname\ outperforms the set point policy (fixed temperature of $25\degree C$).} The distributed policy outperforms the centralized one, as we intuitively expect.

\begin{figure}[htp]
    \centering
        \subfigure[Comparison of policy in centralized control settings: the balanced weights method optimizes for both energy cost and user comfort; taking $w_c = 0$ yields a policy close to ambient temperature; and $w_e = 0$ adapts to user preferences.
        ]{\includegraphics[width=.36\textwidth]{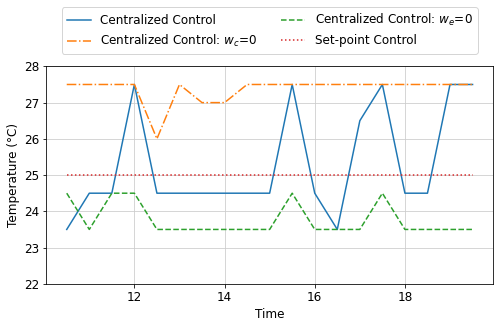}}
        \subfigure[Distributed control policy. The policy personalizes decision-making for each store, given user movements, preferences, and energy costs. 
        ]{\includegraphics[width=.42\textwidth]{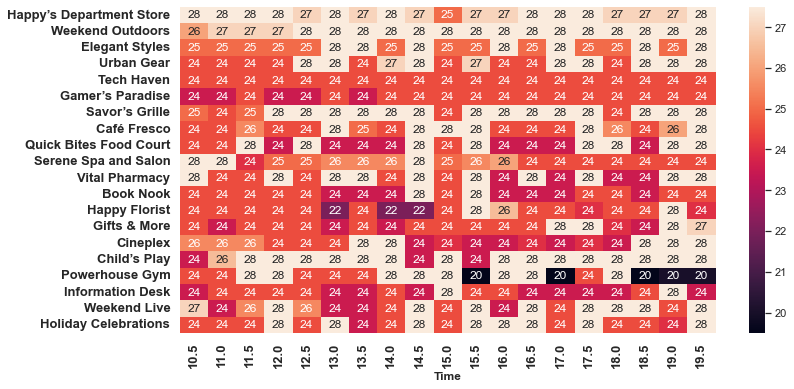}}
        \subfigure[\textcolor{violet}{Baseline}: Distributed control policy \textcolor{violet}{in a energy-focused setting ($w_c=0$).} The policy generates controls 
        to maximize energy savings.]{\includegraphics[width=.42\textwidth]{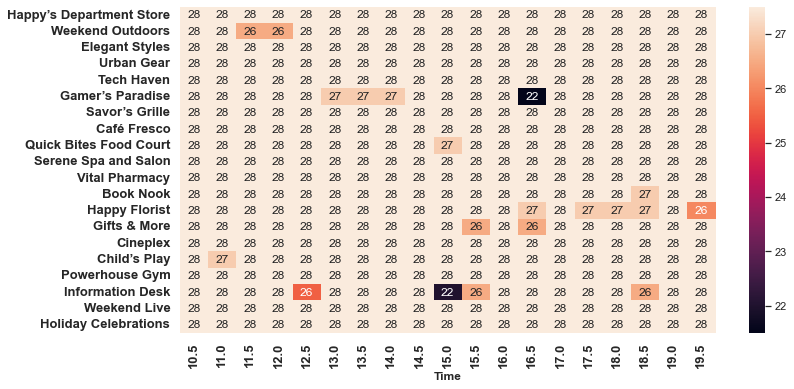}}
        \subfigure[Baseline: Distributed control policy \textcolor{violet}{in a user-comfort-focused setting ($w_e=0$).} The policy adapts temperature controls to maximize user comfort.]{\includegraphics[width=.42\textwidth]{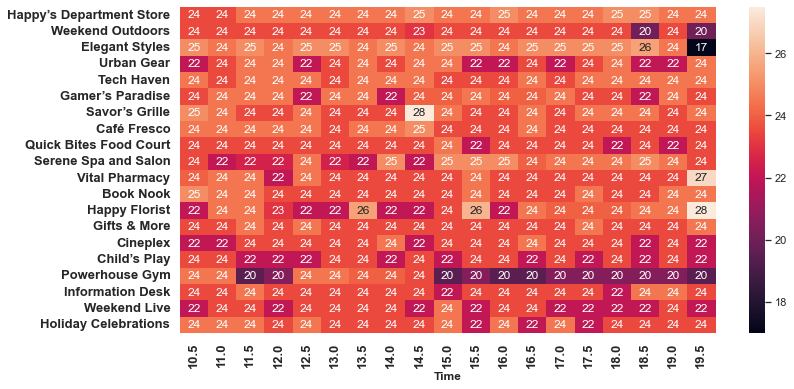}}
    \caption{Comparison of policy (temperature control) in an online scenario using a balanced weights approach. The \algoname\ shows that adaptive control based on user movements and preferences results in more effective optimizations.} 
    \label{fig:policyCompare}
\end{figure}

Figure \ref{fig:policyCompare}.a illustrates the policies for centralized control and set-point control. The centralized control policy adjusts temperatures throughout the day, largely due to varying user activity levels. 
When focusing on user comfort only, the policy sets temperatures significantly lower compared to other settings to accommodate user preferences, but this results in higher energy costs, \textcolor{violet}{and vice versa when focusing on energy costs.} 

Figure \ref{fig:policyCompare}.b displays the distributed control policy. The policy adjusts the temperature for each store based on user activity and group, setting an optimum temperature tailored to smaller groups. This policy allows for personalized temperature decisions, effectively catering to specific needs. During low activity periods, temperatures are set closer to the ambient temperature to minimize energy costs. In Figure \ref{fig:policyCompare}.c, which focuses solely on energy costs, the temperatures are significantly higher, aligning more closely with ambient conditions (30\degree C). Conversely, in Figure \ref{fig:policyCompare}.d, when focusing on user comfort, the policy sets temperatures significantly lower, as intuitively expected.

\section{Discussion and Conclusion}\label{sec:conclusion}
\textcolor{violet}{In this paper, we propose an LLM-based digital twin to simulate human activity to optimize user comfort and energy savings. Our simulator uses the capabilities of LLMs to model multiple population groups and their distinct behavior and preferences in the mall. A reinforcement learning algorithm is then used for temperature control.} LLMs can simulate complex human behaviors, facilitating efficient training of the RL algorithm for user-preference-based \hanqing{CPS controls}. Additionally, distributed control surpasses centralized control in performance, due to its better accommodation of personalized environments. 

\hanqing{Future work could 
focus on making the LLM-based digital twins more accurate simulators of human behaviors, to enable effective learning of \hitl~control policies.} 
For instance, through finetuning them
with occasional human feedback for specific applications, augmenting LLM prompts which generate the simulations or augmenting the LLM-generated data with existing (limited) datasets, or combining the policy trained on the digital twin with one trained on partial human data.
\hanqing{The \algoname~ framework, with its LLM digital twin-based generation of human data, 
is adaptable to other distributed CPS scenarios requiring human interaction, such as traffic management, public transportation planning, etc. For example, the LLM-based digital twin can simulate the mobility and preferences of different population groups across the day in amusement parks and other tourist attractions, and hence optimize attraction and resource placement. 
}

\bibliographystyle{IEEEtran}
\bibliography{references}
\begin{appendices}


\end{appendices}

\end{document}